\documentstyle[aas2pp4,tighten,times]{article}

\newcommand\mathmode[1]{\ifmmode {#1} \else {$#1\mkern-5mu$} \fi}

\def\ki2{$\chi^2$}

\def\b-v{$(B-V)$\kern- .15em$_{\circ}$}     
\def\md{\kern- 0.2em\raise 1.75ex \hbox{$\scriptstyle m$}
    \kern- 1em .}                           
\def\s.{\kern+ .1em\lower 0.5ex\hbox{$\buildrel ^{\prime\prime} \over
    {\rm .} \kern- .38em$}}                 
\def\d.{\kern+ .1em\lower 0.5ex\hbox{$\buildrel ^{\rm d} \over {\rm .}
    \kern- .38em$}}                         
\def\gta{\lower 0.5ex\hbox{$ \buildrel>\over\sim\ $}}
\def\lta{\lower 0.5ex\hbox{$ \buildrel<\over\sim\ $}}
\def\ss.{\kern+ .1em\lower 0.5ex\hbox{$\buildrel {\rm s}
    \over {\rm .} \kern- .05em$}}           
\def\ra#1 #2 #3 #4{$ #1^h#2^m#3\ss. #4$}    
\def\dec#1 #2 #3 {$
  #1^\circ#2\hbox{\char'023}#3\hbox{\char'175}$}    
\def\sec#1 {$ #1\hbox{\char'175}$}

\def\ls{\vskip\the\baselineskip}           

\def\msun{\mathmode{M_\odot}}

\def\gmodes{$g$-modes }
\def\pmodes{$p$-modes }

\def\P0{\Pi_o}
\def\Teffo{T_0}
\def\Teff{T_{\rm eff}}
\def\Menv{M_{\rm env}}
\def\Mcore{M_{\rm core}}
\def\Msun{M_{\rm \odot}}

\def\d2i{{\partial^2 I_\nu\over\partial T^2}\G|_{\Teffo}}

\def\10spaces{\ \ \ \ \ \ \ \ \ \ }
\def\G#1{\Bigg#1}                           

\slugcomment{ Submitted to {\it The Astrophysical Journal (Letters) 1996 May 17}}

\begin{document}

\title{The Potential of Asteroseismology for Hot, Subdwarf B Stars: A
New Class of Pulsating Stars?}
\author{S. Charpinet, G. Fontaine, and P. Brassard}
\affil{D\'epartement de Physique, Universit\'e de Montr\'eal, C.P.
                 6128, Succ. Centre-Ville, Montr\'eal, Qu\'ebec, Canada,
                 H3C 3J7\\ 
                 charpinet, fontaine, brassard@astro.umontreal.ca\\
                 }
\author{and\\}
\author{B. Dorman\footnote{NAS/NRC Resident Research Associate}}
\affil{Laboratory for Astronomy and Solar Physics, NASA/GSFC,
Greenbelt, Maryland 20771\\
Ben.Dorman@gsfc.nasa.gov\\}

\begin{abstract}

We present key sample results of a systematic survey of the pulsation
properties of models of hot B subdwarfs.  We use equilibrium structures
taken from detailed evolutionary sequences of solar
metallicity ($Z$ = 0.02) supplemented by grids of static envelope models
of various metallicities ($Z$ = 0.02, 0.04, 0.06, 0.08, and 0.10).  We
consider all pulsation modes with $l$ = 0, 1, 2, and 3 in the 80--1500 s
period window, the interval currently most suitable for fast photometric
detection techniques.  We establish that significant driving is often
present in hot B subdwarfs and is due to an opacity bump associated with
heavy element ionization.  We find that models with $Z$ $\geq$ 0.04 show
low radial order unstable modes; both radial and nonradial ($p$, $f$, 
and $g$) pulsations are excited.  The unstable models have $\Teff$
$\lta$ 30,000 K, and log $g$ $\lta$ 5.7, depending somewhat 
on the metallicity.  We emphasize
that metal enrichment needs only occur locally in the driving region.  
On this basis, combined with the accepted view that local
enrichments and depletions of metals are common place in the envelopes
of hot B subdwarfs, we predict that some of these stars should show
luminosity variations resulting from pulsational instabilities.

\end{abstract}

\keywords{stars: interiors$-$stars: oscillations$-$subdwarfs}

\section{INTRODUCTION}

In the last twenty years, considerable progress has been made in our
understanding of the physical properties and evolutionary status of hot,
hydrogen-rich subdwarf B stars (see, e.g., \cite{GrS};
\cite{H87}, \cite{H91}; \cite{alet94}; \cite{set94}; \cite{vet95}).  
Following the scenario originally proposed by Heber and
collaborators (e.g., \cite{Het84}), it is currently believed that
subdwarf B stars are $\sim$ 0.5 $\Msun$ objects with hydrogen envelope
masses that are small enough that, after core helium exhaustion on the
extended horizontal branch (EHB), they never evolve toward the
asymptotic giant branch (AGB).  They remain at high
effective temperatures ($\Teff$ $\gta$ 20,000 K) throughout their
core-burning evolution. In the post-HB, He-shell-burning stage of evolution  
the models are referred to as ``AGB-Manqu\'e'' objects (see \cite{gr90}) and are associated with the field subdwarf O stars
(\cite{dor} and references therein).
A number of post-EHB evolutionary studies have found
their way in the literature in the last several years (see \cite{d95} for a review).   On the basis of the evolutionary models of
Dorman, Rood, \& O'Connell 1993, \cite{bet94} have shown that the
subdwarf B stars most likely evolve to become the (relatively rare)
low--mass DAO white dwarfs, thus contributing to a small percentage of
the total white dwarf population.

While asteroseismology is proving to be an extremely powerful tool in
studying other types of stars, its potential has not yet been studied
for subdwarf B stars partly because of the lack (until recently) of
sufficiently realistic equilibrium structures.  Furthermore, and to our
knowledge, luminosity variations caused by pulsational instabilities
have not been reported for these stars\footnote{Note that, in the course
of their Str\"{o}mgren photometric program, \cite{bet84} noted
and subsequently reported the possible variability of the classical
subdwarf B star Feige 108.  However, the equipment available to them as
well as time constraints did not allow the gathering of a light curve of
sufficient quality for time-series analysis.}.  However, pulsation
theorists with a keen eye may have noticed for some time that the
potential for driving pulsation modes appears to exist in the envelopes
of subdwarf B stars.  Indeed, early models of such envelopes (\cite{wc81}; 
\cite{wet82}; \cite{v82}; \cite{gkh85}) as well as their more sophisticated modern counterparts are
all characterized by the presence of a HeII--HeIII convection zone.
By analogy with other types of pulsating stars (whose instabilities are
always driven by one form or another of an opacity mechanism), one might
expect that the opacity bump associated with this partial ionization
zone could also excite pulsation modes in hot B subdwarfs.

Motivated in part by this observation (and see \cite{vh91}), we
undertook a systematic exploration of the potential of asteroseismology
for subdwarf B stars.  The purposes of the present $Letter$ are (1) to
report on the salient results of our investigation, (2) to call
attention to the very real possibility that some subdwarf B stars may
undergo stellar oscillations, and (3) to encourage searches for
luminosity variations in such stars.

\section{COMPUTATIONS AND RESULTS}

The first batch of equilibrium models investigated in this study
consists of full stellar models taken from 5 distinct evolutionary
sequences.  These were chosen in order to map a significant fraction of
the region of the $\Teff$--log $g$ plane actually occupied by the known
subdwarf B stars (see \cite{set94} and references therein).  In
this particular region, we retained about 25 models per sequence.  The
evolutionary models were computed with the methods  described in 
Dorman (1992a,b) and \cite{dro}.  The new models use 
the OPAL opacities described by \cite{opal} computed 
in December 1993, which adopted the
 element mix referred to as `Grevesse \& Noels 1993'. Where necessary (during 
He-flashes), we used new low temperature opacities by D. R. Alexander 
(1995, private communication, described in \cite{af94}) 
which were computed with the same element mix. 
These smoothly match the OPAL opacity set within
the hydrogen ionization zone. The other difference in the physics
was the use of the Itoh et al. (1983; 1993a,b; 1994a,b) conductive
opacities.

Each sequence describes the evolution of an AGB-Manqu\'e star with a core
mass of 0.4758 \msun.  The sequences differ in that different envelope
masses are considered: $\Menv =$ 0.0002, 0.0012, 0.0022, 0.0032,
and 0.0042 \msun.  The composition of the envelopes is assumed solar
($X = 0.70388$, $Z = 0.01718$).  The luminosity of each model is provided by
He burning confined to a very small core.  Residual H burning at the
base of the H-rich envelope contributes negligibly to the luminosity
until the AGB-Manqu\'e phase of the evolution.

We first carried out a detailed adiabatic survey of the evolutionary
models with the help of the Galerkin finite--element code of \cite{bet92}.  
Specifically, for each equilibrium model considered, we
computed pulsation periods in the adiabatic approximation for all modes
with $l$ = 0, 1, 2, and 3 in the 80--1500 s period window.  This
interval corresponds to the range of periods most easily detectable with
present--day fast photometric techniques.  Moreover, it is well known
that surface cancellation effects render detection of luminosity
variations excessively difficult for $l \geq 4$ in compact stars
(\cite{dz77}; \cite{rkn82}; \cite{br87}; 
\cite{bfw95}), so such modes are not of direct
observational interest.  We will report elsewhere the
detailed results of this extensive survey, including discussions of the
effects of changing model parameters, the period evolution, the rates of
period changes, and the effects of mode trapping and mode confinement
caused by the composition interfaces in our doubly stratified
equilibrium models.  For the needs of the present paper, in order to
give an idea of the modes to be expected in the 80--1500 s period
window, we simply report on some sample results.

\begin{planotable}{lcccccccc}
\tablewidth{40pc}
\tablecaption{Pulsation Periods for a Typical Subdwarf B Star Model}
\label{tb:periods}
\tablehead{
\colhead{$k$} & 
\colhead{$l=0$} &
\colhead{$l=1$ $(p)$} & 
\colhead{$l=1$ $(g)$} & 
\colhead{$l=2$ $(f,p)$} & 
\colhead{$l=2$ $(g)$} & 
\colhead{$l=3$ $(f,p)$} & 
\colhead{$l=3$ $(g)$}}

\startdata 
0 & 384.13 & ... & ... & 277.45 & ... & 270.11 & ...\nl
1 & 281.70 & 379.97 & 399.07 & 257.90 & 370.99 & 222.67 & 306.83\nl
2 & 226.74 & 279.72 & 669.19 & 223.56 & 407.72 & 205.82 & 355.99\nl
3 & 191.25 & 225.28 & 965.35 & 192.39 & 576.19 & 187.56 & 428.29\nl
4 & 173.54 & 188.71 & 1136.81 & 179.41 & 675.85 & 169.34 & 506.07\nl
5 & 157.05 & 163.60 & 1311.82 & 157.95 & 774.48 & 153.11 & 563.83\nl
6 & 138.66 & 147.90 & ... & 139.38 & 897.80 & 135.79 & 655.69\nl
7 & 114.61 & 134.07 & ... & 126.07 & 1004.21 & 121.56 & 723.91\nl
8 & 105.62 & 120.33 & ... & 115.90 & 1182.10 & 110.75 & 846.59\nl
9 & 96.96 & 108.84 & ... & 106.04 & 1361.08 & 102.54 & 975.36\nl
10 & 89.50 & 99.99 & ... & 97.15 & 1429.47 & 95.03 & 1030.05\nl
11 & 83.38 & 93.28 & ... & 89.77 & ... & 88.03 & 1130.15\nl
12 & ... & 87.18 & ... & 83.87 & ... & 81.88 & 1259.89\nl
13 & ... & 81.28 & ... & ... & ... & ... & 1337.26\nl
14 & ... & ... & ... & ... & ... & ... & 1425.84\nl
\enddata

\end{planotable}

Table~\ref{tb:periods} lists the pulsation periods for a typical model of a hot B
subdwarf. This model belongs to the sequence with $\Mcore$ = 0.4758
$\Msun$ and $\Menv$ = 0.0012 $\Msun$.  It has an age of $\sim 8.4 \times
10^{7}$ yrs (time elapsed since the zero age HB phase), a surface
gravity log $g$ = 5.46, and an effective temperature $\Teff$ = 27,500 K.
The table gives the pulsation period (expressed in seconds) as a
function of the radial order $k$.  We distinguish between the radial
modes ($l$ = 0) and the $p$ and $g$ branches for nonradial modes with
$l$ = 1, 2, and 3.  Our results indicate that subdwarf B stars have rich
period spectra, easily accessible with current--day observational
techniques (white light fast photometry, for example).  Provided we can
demonstrate that some of these modes can be excited, this should
motivate observational searches for pulsational instabilities in these
stars.

In the second step of our investigation, we therefore carried out a
stability analysis of our equilibrium models with the help of the
finite--element nonadiabatic pulsation code briefly described in \cite{fet94} and \cite{bfb96}.  In order to
understand the results and be able to identify the regions of driving
(if any), we considered the variations with depth of the thermal
timescale and of the so--called work integral.  Figure 1 illustrates a
typical case and refers to the equilibrium model whose periods are
provided in Table~\ref{tb:periods}.  The solid curve in the upper panel shows how the
local thermal timescale, $\tau_{\rm th}$ (expressed in seconds), varies as
a function of the fractional mass depth, $\log q \equiv \log (1-M(r)/M)$,
in this particular unperturbed model.  As a comparison, the values of
the periods of the $p(k=3)$, $f(k=0)$, and $g(k=3)$ modes with $l=2$ are
indicated.  In addition, the locations of the He burning core, the H
burning shell, and the HeII--HeIII convection zone are also shown.
Because the local thermal timescale in the thermonuclear burning regions
is orders of magnitude larger than the periods of interest (in addition
to the fact that the H burning shell is extremely weak), it is clear
that those regions cannot drive efficiently the pulsation modes.  In
practice, the $\epsilon$ mechanism thus appears to be irrelevant for
exciting pulsation modes in hot B subdwarfs.  In contrast, the local
thermal timescale is of the same order of magnitude as the interesting
pulsation periods in the envelope convection zone, which suggests, as
anticipated above, that driving could be efficient in that zone.

A look at the lower panel in the figure reveals, however, a small
surprise and an important result.  The solid curve refers to the
integrand of the work integral for the $l=2$, $f$ mode mentioned above.
(We emphasize that, while specific to this particular mode, the results
shown here are also typical of all other modes of interest.)  The values
of $dW/dr$ are positive in driving regions and negative in damping
regions.  We find that driving is negligible in the convection zone.  
This constitutes the small surprise.
In retrospect, it appears that the convection zone is located too high
in the envelope, in a region that contains very little mass and,
therefore, carries practically no weight in terms of driving (or
damping).  The important result is that the main driving region in hot B
subdwarfs is associated with an opacity bump due to heavy element
ionization.  This is shown by the dashed curve which illustrates the run
of the Rosseland opacity in the unperturbed model.  As expected, the
opacity shows a maximum in the convection zone due to HeII--HeIII
partial ionization, but it also shows a secondary maximum around $\log q
\simeq -9.2$, obviously associated with the driving region, and which
tends to disappear when the heavy element content (assumed to be $Z \sim
0.02$ in this model) is decreased.  While the local thermal timescale in
the driving region is larger than the periods of interest (see upper
panel), our results indicate that it is not so much larger as to
render the coupling totally inefficient.  Moreover, the eigenfunction
amplitudes are relatively large in the driving region.  The realization
that driving is related to the metallicity provides an important clue as
to the possible existence of real subdwarf B pulsators.

The specific mode considered in the lower panel of Figure 1
turns out to be stable; the damping region below the driving region
contributing somewhat more to the overall work integral.  This result is
quite typical of all of the  evolutionary models computed with $Z \sim
0.02.$  Indeed, in our detailed nonadiabatic survey of those equilibrium
models, we found no unstable mode.  However, in many cases, we found
that the overall damping rates 
were very small, suggesting to us that such modes
could perhaps still be excited in real stars (whose detailed structure
might, of course, be different from that of the particular models we
considered).  In view of the correlation we uncovered between driving
and metallicity, and the fact that local variations of metallicity $are$
expected in subdwarf B stars (see below), it seemed natural to
investigate the effects of changing the metallicity.

In the final step of our analysis, we thus constructed a second batch of
equilibrium models taking into account variations of $Z$ in the H-rich
envelope.  Contrary to our previous models, those are static (i.e.,
non-evolving) structures.  The models were built with a version of
the stellar code described by \cite{bf94} and Brassard
et al. (1996) suitably modified to produce envelope structures extending
as deep as $\log q = -0.05$.  The constitutive physics used is nearly
identical to that used previously in the construction of our full
evolutionary models, apart from the use of slightly newer OPAL
opacities computed in 1995.  Tests indicate that our ignoring of the small He
burning core increases somewhat the values of the periods of the \gmodes
(by 10-20 \%, typically), but does not affect those of the \pmodes 
which, contrary to the former, are formed
essentially in the envelope region.  We will provide elsewhere 
more details on these models.

We constructed 5 different model grids, one each for a fixed metal
content $Z$ = 0.02, 0.04, 0.06, 0.08, and 0.10 in the H-rich envelope
($X$ = 0.70).  Each grid consists of 72 unperturbed models covering a
rectangular region in the $\Teff$--log $g$ plane overlapping with the
region where hot B subdwarfs are actually found ($4.34 \leq \log \Teff
\leq 4.62$ in steps of 0.04, and $4.8 \leq \log g \leq 6.4$ in steps of
0.2).  In these exploratory calculations, we fixed the total mass of
each model to 0.48 $\Msun$ and the H-rich envelope mass to $\log q(H) =
-4$.  Tests indicate that stability does not depend on the value of
$\log q(H)$.

{\it In agreement with our expectations concerning the effects of
metallicity, we found unstable pulsation modes for
models with $Z \geq 0.04$.}  For these models, there is a blue edge
temperature located somewhere between $\log \Teff$ = 4.46 and 4.50, and
which shows a weak sensitivity to metallicity.  We
tentatively interpret this as the consequence of the fact that
metallicity affects primarily the size of the opacity bump (and, hence, the
magnitude of the driving), but not strongly its location in a star of a given
effective temperature. The blue edge itself must be due to the outward
displacement of the driving region to the outermost layers (which carry
little weight in the overall stability of a mode) as the effective
temperature increases.  Note that our actual value of the blue edge
temperature ($\Teff \sim 30,000$ K) may also depend on the assumed mass of
the models (currently 0.48 \msun), but this remains to be
investigated.  We find also that stability does depend on the assumed
surface gravity.  Unstable models have log $g \lta$ 5.7, with a weak
dependence on metallicity, but this remains also to be investigated in
more detail.

\begin{planotable}{lcccccc}
\tablewidth{40pc}
\tablecaption{Unstable Modes in Typical Subdwarf B Star Models}
\label{tb:unstable}
\tablehead{
\colhead{$l$} & 
\colhead{$k$} &
\colhead{$P$(s), $Z$=0.06} & 
\colhead{$\tau_{\rm e}$(yrs), $Z$=0.06} & 
\colhead{$P$(s), $Z$=0.10} & 
\colhead{$\tau_{\rm e}$(yrs), $Z$=0.10}}

\startdata 
0 & 1 & 427.2 & stable & 432.0 & $7.9 \times 10^{-2}$\nl
0 & 0 & 580.6 & $2.2 \times 10^{0}$ & 588.5 & $4.4 \times 10^{-1}$\nl
\nl
1 & 2 (p) & 425.1 & stable & 429.9 & $8.5 \times 10^{-2}$\nl
1 & 1 (p) & 478.6 & $2.2 \times 10^{1}$ & 461.0 & $3.6 \times 10^{0}$\nl
1 & 1 (g) & 849.7 & $7.6 \times 10^{4}$ & 816.7 & $1.3 \times 10^{4}$\nl
\nl
2 & 1 (p) & 424.0 & stable & 428.1 & $9.3 \times 10^{-2}$\nl
2 & 0 (f) & 533.9 & $8.0 \times 10^{1}$ & 513 9 & $7.1 \times 10^{0}$\nl
2 & 1 (g) & 573.8 & $2.4 \times 10^{0}$ & 579.4 & $4.1 \times
10^{-1}$\nl
2 & 2 (g) & 802.0 & $1.2 \times 10^{4}$ & 770.6 & $2.4 \times 10^{3}$\nl
\nl
3 & 0 (f) & 432.1 & stable & 428.2 & $1.2 \times 10^{-1}$\nl
3 & 1 (g) & 562.9 & $2.7 \times 10^{0}$ & 564.7 & $4.4 \times
10^{-1}$\nl
3 & 2 (g) & 605.8 & $7.0 \times 10^{1}$ & 584.6 & $2.2 \times 10^{0}$\nl
3 & 3 (g) & 795.0 & $3.5 \times 10^{3}$ & 764.2 & $7.8 \times 10^{2}$\nl
3 & 4 (g) & 980.6 & $8.3 \times 10^{3}$ & 942.3 & $1.7 \times 10^{3}$\nl
3 & 5 (g) & 1147.2 & stable & 1095.4 & $1.5 \times 10^{3}$\nl
\nl
\enddata

\end{planotable}

We end this section by presenting, in Table~\ref{tb:unstable}, typical examples of
unstable modes.  We list there the periods and e-folding times of the
unstable modes found in 2 models, one computed with $Z$ = 0.06, and the
other with $Z$ = 0.10.  Both have log $g$ = 5.2 and log $\Teff$ = 4.42.
For given modes, the periods are significantly larger than in
Table~\ref{tb:periods}
because of the lower surface gravity considered here (although for the
\gmodes the differences are partly due also to our use of envelope
models instead of full stellar models).  As
expected, the driving is clearly more efficient in the high--metallicity
model; a wider band of modes are excited in this particular model as
compared to its $Z$ = 0.06 counterpart.  In both cases, the excited
modes have low values of the radial order $k$, and radial as well as
nonradial pulsations are predicted.  Since the e-folding times are
generally significantly shorter than the typical evolutionary time of a
hot B subdwarf ($\sim 10^{8}$ yrs), these linear instabilities would
normally develop into observable amplitudes.

\section{DISCUSSION AND CONCLUSION}

The results presented in this paper pave the way to the exciting
possibility of being able to use asteroseismology as a probe of the
internal constitution of subdwarf B stars.  A subgroup of these stars
may indeed constitute a new class of pulsating stars.  Of course, this
possibility hinges here on the question of metal enrichment in the
driving region.  We emphasize the fact that this enrichment needs only
occur in the driving region {\it itself} and not necessarily everywhere in
the envelope (as we assumed in the relatively crude envelope models used
in the present study).  That heavy elements can show local enrichments
in certain regions of a hot B subdwarf envelope, and local depletions in
others, is not only plausible but is also {\it expected}.  Indeed, subdwarf
B stars all show peculiar surface abundances whose study constitutes a
most active and interesting subfield of stellar physics in its own
right.  It is currently believed that the abundance anomalies observed
in the atmospheres of these stars result from the competition between
gravitational settling, radiative levitation, and weak stellar winds
(\cite{miet85}, \cite{miet89}; \cite{let85}, \cite{let87}; \cite{H87}, 
\cite{H91}; \cite{bet88}; \cite{cet96}).  In the most
recent investigation, \cite{cet96} have carried out new 
computations of radiative levitation on metals in models of hot B
subdwarfs using the same tools considered previously in the white dwarf
context by \cite{cfw95} and \cite{cet95}.  Of high relevance here, they
found that metals, and in particular iron (a major contributor to the
opacity), do levitate in the envelopes of these stars.  Local
overabundances of more than an order of magnitude can be built through
this process.  In the case of iron, the maximum overabundance occurs
near $\log q \simeq -9,-10$.  This result greatly enhances the plausibility
that driving occurs as a result of a local enhancement of the metal
content\footnote{A similar phenemenon may be responsible for mode
excitation in rapidly oscillating Ap stars.}.  
Of course, a detailed study of the driving process in hot B
subdwarfs will ultimately require the use of more detailed envelope
structures taking into account diffusion processes.  {\it In the
meantime, we feel confident enough to risk the prediction that some
subdwarf B stars should show luminosity variations resulting from
pulsational instabilities.}  We believe that the material presented in
this paper warrants further theoretical investigations of the pulsation
properties of these stars and, above all, searches for luminosity
variations.  We have undertaken both.

\acknowledgements

This work was supported in part by the NSREC of Canada 
and by the fund FCAR (Qu\'ebec). We thank Matt Wood 
for providing the original versions of the conductive 
opacity subroutines. B. D. acknowledges partial support
from NASA RTOP 188-41-51-03.

\onecolumn

\begin{figure}[h]
\vskip5.5truein
\includegraphics{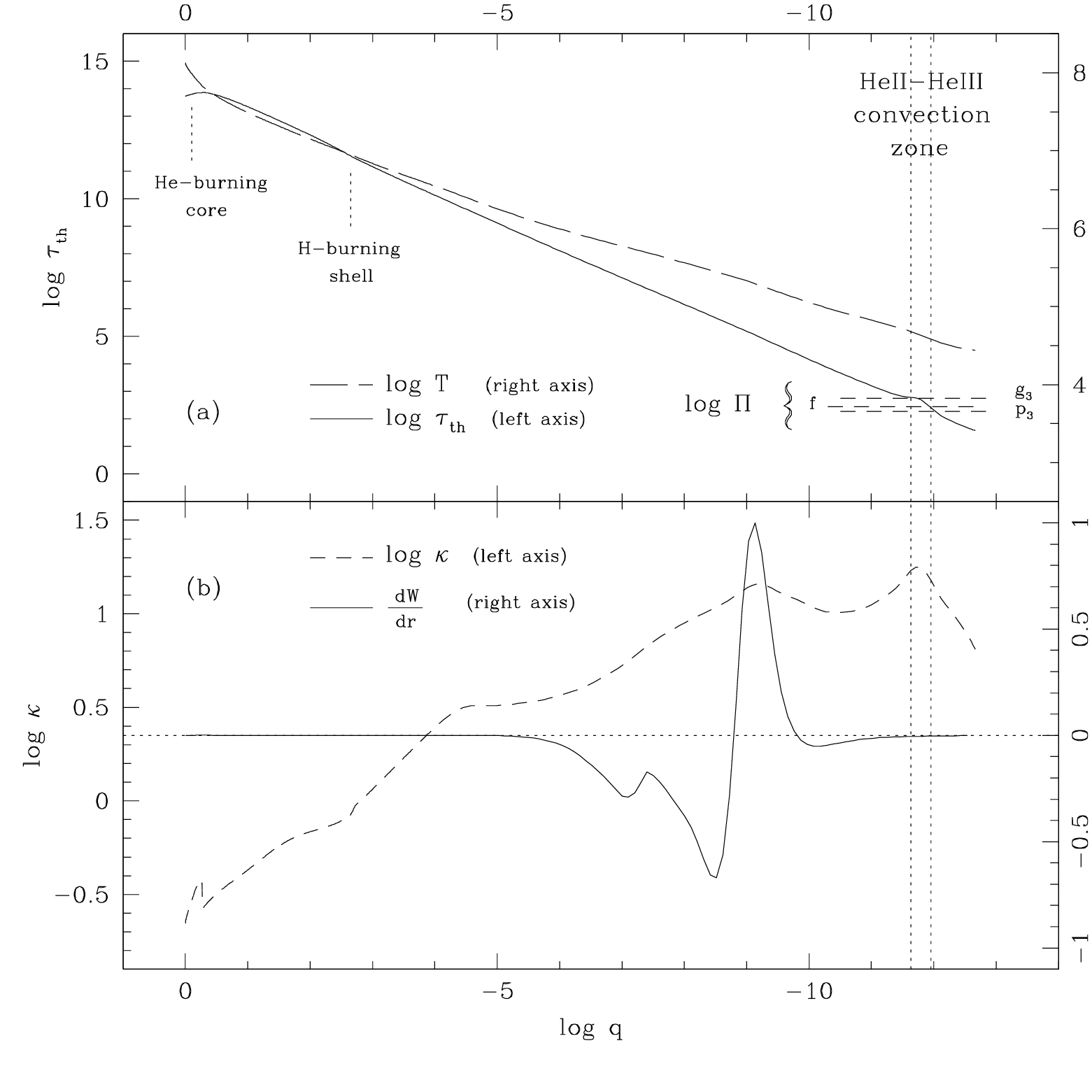}
\caption{
(a) Run of the thermal timescale (expressed in seconds) $vs$
fractional mass depth in a typical subdwarf B model (solid curve).  The
locations of the He burning core, the residual H burning shell, and the
HeII-HeIII convection zone are indicated.  As a comparison, the values
of the periods of a few low radial order modes with $l$ = 2 are shown.
The temperature distribution (long-dashed curve) is also illustrated.  
(b) Run of the Rosseland opacity (dashed curve) and of the integrand of
the work integral for the $f$ mode with $l$ = 2 (solid curve).  The driving
region (positive values of $dW/dr$) is clearly associated with an
opacity bump, itself caused by heavy element ionization.
}
\end{figure}

\twocolumn


\end{document}